\documentclass[a4paper,twoside]{article}
\newcommand{\affil}[1]{$^{\rm #1}$}
\usepackage[authoryear]{natbib}
\bibpunct{(}{)}{;}{a}{}{,}
\usepackage{graphicx}
\date{} 
\newcommand{\kms}{\mbox{km\,s$^{-1}$}}
\newcommand{\mH}{m_{\scriptscriptstyle \rm H}}
\title{\large\bf\flushleft Compton Scattering of Fe K$\alpha$ Lines from Accreting White Dwarfs}
\author{\parbox{\textwidth}{\flushleft
\vspace{-0.5cm}
{\it Zdenka Kuncic\affil{1,4}, Kinwah Wu\affil{2} and Jason G. Cullen\affil{3}}\\
\vspace{0.4cm}
{\small \affil{1}\,School of Physics, University of Sydney, NSW 2006, Australia}\\ 
{\small \affil{2}\,Mullard Space Science Laboratory, University College London, Holmbury St Mary,  
     Dorking, Surrey, RH5~6NT, United Kingdom}\\ 
{\small \affil{3}\,ADI Ltd., Canberra, ACT 2612, Australia} \\ 
{\small \affil{4}\,Email: z.kuncic@physics.usyd.edu.au}
}}
\begin{document}
\twocolumn[
\begin{changemargin}{.8cm}{.5cm}
\begin{minipage}{.9\textwidth}
\vspace{-1cm}
\maketitle
%
%
\small{\bf Abstract:}
Compton scattering in the bulk accretion flow of the accretion column 
  in magnetic cataclysmic variables (mCVs) 
  can significantly shift photon energies in the X-ray emission lines  
  resulting  from  accretion shocks. 
In particular, Compton recoil can potentially broaden the 6.7 and 6.97~keV
Fe K$\alpha$ emission lines 
  produced in the post-shock region, 
  and contaminate the fluorescent 6.4~keV neutral Fe K$\alpha$ line 
  reflected off  the white dwarf surface.
We present nonlinear Monte Carlo simulations 
  demonstrating these effects, 
  and we discuss the interpretation of observed Fe K$\alpha$ linewidths 
  in mCVs in light of these new results.
The implications for other accreting compact objects are also discussed.


\medskip
\medskip
\end{minipage}
\end{changemargin}
]
\small

\section{Introduction}  

Magnetic cataclysmic variables (mCVs) 
  are low-mass close binaries with a Roche-lobe overfilling red dwarf  
  transferring material to a magnetic white dwarf 
  (see e.g.\ Warner 1995; Cropper 1990).  
The two main subclasses of mCVs are the polars (AM Herculis systems)  
  and the intermediate polars (IPs).  
In a polar, the white dwarf has a sufficiently strong magnetic field 
  ($B \sim 10 - 200$~MG)
  such that the white dwarf and its companion star are locked into synchronous rotation.  
The magnetic field also prevents the formation of an accretion disk, 
  causing the material from the red dwarf to be accreted onto the white dwarf 
  directly along its magnetic poles.  
Although the magnetic field of a white dwarf in an IP is weaker, 
  allowing an accretion disk to form,  
 the inner part of the disk is truncated by the white-dwarf magnetic field.  
In both polars and IPs, 
  the accretion flows onto their white dwarfs are confined by the magnetic field,  
  and accretion columns are formed, 
  channelling the accreting matter to the white-dwarf magnetic polar regions.  

An accretion shock often develops in the accretion column 
  such that the supersonic accretion flow becomes subsonic in 
the downstream, post-shock region, before settling onto the white-dwarf surface.
The shock heats the accreting matter to temperatures 
 \begin{equation}   
  kT_{\rm s}  \approx  
   {3 \over 8} {{G M_{\rm w} \mu \mH}  \over {(R_{\rm w} + x_{\rm s})}}  
\qquad ,
\label{e:Ts} 
\end{equation} 
  where $M_{\rm  w}$ and $R_{\rm w}$ are the white-dwarf mass and radius, 
  $\mu $ is mean molecular weight, and $x_{\rm s}$ is the shock height,
  which is typically $\ll R_{\rm w}$ 
  (see, however, Cropper et al.\ 1999 for cases with large shock heights).    
For white-dwarf masses of $0.5-1.0 M_\odot$ and typical mCV parameters,  
  the shock temperature is
  $kT_{\rm s} \sim 10-40\,$keV ($\sim 1 - 4 \times 10^8$~K).   
The hot plasma in the post-shock region is cooled by emitting 
  bremsstrahlung X-rays and optical/infra-red cyclotron radiation  
  (Lamb \& Masters 1979; Wu 2000 and references therein).   

At a plasma temperature of $kT \sim 1 - 50$~keV, 
  heavy elements such as Iron (Fe) are highly ionised, 
  resulting in an abundant population of H- and He-like ions
  (see Fujimoto 1998 and Wu, Cropper \& Ramsay 2001 for the ionisation structure 
  of post-shock flow in mCVs). 
The 6.97~keV Lyman-$\alpha$ line is emitted 
  due to the $2p~^2P \rightarrow 1s~^2S$ transition in the H-like Fe~XXVI ions. 
The 6.70~keV He4 line is emitted from the He-like Fe~XXV ions 
  for the $2p~^1P_1 \rightarrow 1s~^1S_0$ transition; 
  and the 6.67 and the 6.68~keV He5 lines, 
  for the $2p~^3P_1 \rightarrow 1s~^1S_0$ 
  and the $2p~^3P_2 \rightarrow 1s~^1S_0$ transitions respectively. 
The 6.64~keV He6 line, 
  due to the $2p~^3S_1 \rightarrow 1s~^1S_0$ forbidden transition, 
  is generally suppressed in the shock-heated regions of mCVs,   
  because of a high ($> 10^{14}~{\rm cm}^{-3}$) electron number density.  

The natural widths of these Fe~K$\alpha$ lines are small. 
Realistically, the lines will be Doppler broadened by 
  the bulk motion of the emitters in the flow,
  and also by scattering processes.     
X-ray observations of mCVs have indeed revealed substantially broadened
Fe K$\alpha$ emission lines \citep{Hellier98,HelMuk04}, 
  though flow-velocity induced Doppler shifts are yet to be verified.  
The line photons may undergo a large number of resonant scatterings with the ions, 
  but this requires the scattering region to have small velocity gradients 
(See Terada et al.\ 2001 for a detail study of resonant scattering in mCVs).  
Electron (Compton) scattering can shift line photon frequencies significantly 
  and thereby allow the photons to escape resonance trapping.  
It also can cause substantial line broadening.

A previous study (Wu 1999) proposed 
  that cold electrons in the upstream pre-shock bulk flow of mCVs
  can scatter a significant number of line photons emitted from the shock-heated region, 
  and any 6.4~keV fluorescent Fe line from the white-dwarf surface  
  could be contaminated by the down-scattered 6.7~keV He-like Fe line photons.  
A recent study (Matt 2004) showed that such scattering could be important 
  in that it modifies the X-ray polarization properties.   
In this paper, we investigate the effects of electron (Compton) scattering
  in the accretion column of mCVs by means of numerical simulations. 
We focus on its effects on the H-like, the He-like, and the neutral
Fe K$\alpha$ lines.
We employ a non-linear Monte Carlo algorithm  (Cullen 2001a,b), 
   which improves on other similar codes \citep{Stern95,Hua97} 
   by taking into account nonuniformity in the velocity, density and temperature profiles 
   in the scattering region. 
The paper is organised as follows:
   in \S 2, we briefly discuss various line broadening mechanisms;  
   in \S 3 we outline the geometry and physical parameters of our model 
     for the line-emitting and Compton scattering regions in mCVs, and 
    we summarise the Monte Carlo method used in our simulations; 
   results are presented and discussed in \S 4; 
   and conclusions are given in \S 5.

\section{Broadening of Fe~K$\alpha$ Lines from mCVs} 

The post-shock region of mCVs are mostly 
optically thin to bremsstrahlung X-rays,   
  and the coronal approximation is usually applicable   
  for line emissivity  calculations 
  (e.g.\ Fujimoto 1998; Wu, Cropper \& Ramsay 2001).   
The thermal Doppler widths of the Fe K$\alpha$ H-like line are about 5~eV 
  at temperatures $\sim 10$~keV,   
  and the natural width of the line is about 1~eV. 
The Fe lines can, however, be broadened by the velocity dispersion 
  of the accretion flow in the emission region.   
The infalling material will hit the accretion shock with speeds
$\sim 6000\, \kms$; if the shock is adiabatic, the post-shock speed will be
$\sim 4$ times slower. 
Thus, the velocity dispersion of emission lines in an orbital-phase averaged spectrum 
could reach up to $1000\, \kms$,  corresponding to a broadening of about 20~eV. 

Distortion of the line profiles can also be caused by electron (Compton) scattering.  
A line photon can gain energy through collisions with hot electrons 
  (i.e., $kT_{\rm e} > E_{\rm c}$, where $T_{\rm e}$ is the electron temperature 
  and $E_{\rm c}$ the line centre energy), 
  and the fractional energy change per scattering is \citep{Pozdnyakov77}
\begin{equation} 
 { {\Delta E}  \over  E} \ 
  \approx \  \sqrt{ {{2 kT_{\rm e}} \over { {m_{\rm e} c^2}}}}  \ 
  \approx \  0.064~\left({{{kT_{\rm e}}\over  {1~{\rm keV}}}}\right)^{1/2} \  . 
\end{equation}  
For cold electrons ($kT_{\rm e} \ll E_{\rm c}$), the photon loses energy 
  due to recoils. 
For a He-like Fe line, the energy shift per scattering is  
\begin{equation}
  {{\Delta E} \over E} \    \approx   \  {E \over {m_{\rm e} c^2}} \  
      \approx \  0.014~\left({E \over {{\rm 6.7~keV}}} \right) \ . 
\end{equation}     
Although the energy shift per scattering is small,    
  the Fe line can be broadened significantly after multiple scatterings.   

For an mCV with a specific accretion rate 
  of $\sim 1-10 \, {\rm g \, cm}^{-2}{\rm s}^{-1}$,  
  the thickness of the shock-heated region is 
  $x_{\rm s} \sim 10^7$~cm,  
  and the electron number density is $n_{\rm e} \sim 10^{16}~{\rm cm}^{-3}$. 
The corresponding Thompson optical depth is $\tau \sim 0.1$. 
The mean number of scatterings is $N\approx$~max($\tau^2,\tau$) 
  (Rybicki \& Lightman 1979). 
Thus, one in every ten line photons would encounter an electron  
  before leaving the post-shock region. 
The effective Doppler broadening due to Compton scattering 
  by hot electrons in this downstream region  
  is about $5$\% of the line-centre energy. 
  
Compton scattering by the cold, upstream electrons 
  in the pre-shock region
  can, on the other hand,  cause much more substantial line broadening. 
Despite the fact that the electron number density in the upstream flow 
  could be an order of magnitude lower than that in the downstream, post-shock
region, 
  the linear size of the accretion column between the accretion shock 
  and the magnetospheric coupling region could be more than two orders of magnitude
  larger ($\sim 10^9$~cm) than the thickness of the shock-heated region.  
The effective line-of-sight electron scattering optical depth $\tau$ is therefore
anisotropic,  
  with values larger than unity at certain viewing angles.  
We would therefore expect observable Compton recoil signatures 
  in the X-ray emission line spectra of mCVs. 
  
Note that line broadening due to resonance scattering 
  is a factor of $(m_{\rm e}/m_{\rm i})^{1/2}$ smaller than 
broadening due to Compton scattering. 
For the Fe H- and He-like lines, 
  this factor is $\sim 1/320$ (Pozdnyakov et al. 1977). 

\begin{figure}
\centerline{\includegraphics[width=8.0truecm]{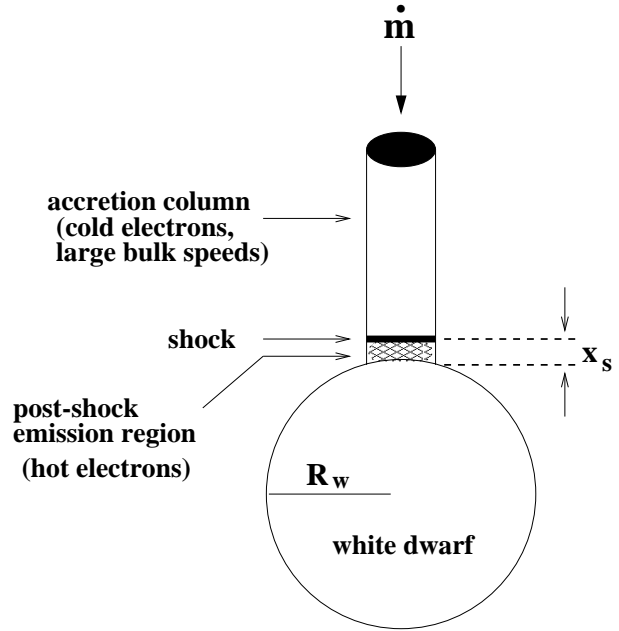}}
\caption{Schematic illustration of the geometry used to model
accretion onto a white dwarf, line emission by shock-heated
electrons, and Compton scattering by both hot and cold electrons
in the accretion column.}
\end{figure}

\section{Modelling Compton Scattering in the Accretion Column} 

\subsection{Line emission and scattering regions}  

We investigate Compton scattering effects on Fe K$\alpha$ lines in mCV accretion
columns using Monte Carlo simulations.
We use line centre energies of 6.4 keV, 6.675 keV, and 6.97 keV, corresponding to
transitions of neutral, He-like, and H-like Fe, respectively.
The neutral and Lyman-$\alpha$ transitions are actually doublets,
with energies 6.391/6.404~keV and 6.952/6.973~keV and branching ratio 1:2, and the He-like
transition has both resonant and intercombination components, and dielectronic 
satellite lines are also associated with the H-like and He-like lines.
These subcomponents remain unresolvable with current X-ray instruments,
and in this work, we simply use the approximate line centre energies in order
to demonstrate the importance of Compton scattering effects on the line profiles.

In the simulations, 
   the line photons are emitted and scattered in the accretion column, 
   which is assumed to be a cylinder with a circular cross section.  
The cyclinder is divided into a geometrically-thin, downstream shock-heated region, 
   and a semi-infinite, upstream pre-shock region. 
The radius of the accretion cylinder is assumed to be 0.1~$R_{\rm w}$.  
The line photons are emitted in the shock-heated region,  
  and scattering occurs in both the post- and pre-shock regions.   
The geometry of the model is shown in Figure~1.
We note that the X-ray emission from the post-shock accretion column illuminates the
white dwarf atmosphere, which could cause Fe~K$\alpha$ fluorescence and additional
Compton recoil effects.
Such effects can be modelled using a Monte-Carlo technique.
In this study, however, we focus only on the scattering of the lines emitted
from the shock-heated plasma.
Our future study will address this issue properly.

The post-shock accretion flow is cooled 
  by emitting bremsstrahlung X-rays and cyclotron optical/infra-red radiation. 
Line cooling and Compton cooling are generally unimportant in most of the post-shock
region and are thus neglected in this study. 
The velocity, density and temperature profiles   
  in the shock-heated region are assumed to be nonuniform, in contrast to previous studies, 
  and are calculated 
  using the hydrodynamic model (with bremsstrahlung and cyclotron cooling)  
  described in Wu, Chanmugam \& Shaviv (1994).
Although resonant scattering in the post-shock region can be important,
we neglect this process in the simple model presented here, although resonance line
trapping is expected to enhance the overall effects of Compton scattering.

We note that the ionisation structure and the line emissivities in the accretion column 
  can be calculated directly \citep{Wu01}. 
In this study, however, we simply assume that 
  the H-like and He-like lines are emitted from a particular stratum in the column
  and the 6.4 keV Fe line arises from the bottom stratum just above the white-dwarf surface. 
A more detailed study of scattering effects 
  with a more self-consistent model of the spatial variation 
  in line emissivities and line-of-sight viewing angles will be presented elsewhere
  (Kuncic et al., in preparation).   
The upstream pre-shock region is assumed to be filled with cold ($kT_{\rm e} \ll 1$~keV)
plasma accreting at the local free-fall speed
$v_{\rm ff} = [GM_{\rm w}/(R_{\rm w}+z)]^{1/2}$, with  densities determined by the
accretion rate $\dot m$, \textit{viz.} $n_{\rm e} = \dot m / (\mu \mH v_{\rm ff})$.   

The parameters that determine the structure of the post-shock region 
   are therefore the white-dwarf mass $M_{\rm w}$, 
   the white-dwarf radius $R_{\rm w}$, the specific accretion rate $\dot m$, 
   and the ratio of the efficiency of cyclotron cooling to bremsstrahlung cooling 
   $\epsilon_{\rm s}$ 
   (see Wu et al. 1994; Saxton, Wu, \& Pongracic 1997). 
While the white-dwarf mass is a free input parameter, 
   the white dwarf radius is obtained 
   from the mass-radius relation given in Koester (1987).  

\subsection{Monte Carlo algorithm}   

Line photons are injected in a stratum, 
  and the propagation vector of the photons are in random directions
  in the rest frame of the flow. 
For each photon, 
  the initial propagation vector is Lorentz transformed 
  out of the rest frame of the emitter  
  into the rest frame of the white dwarf. 
We then determine a (tentative) scattering point, 
  the distance to which is determined 
  using a non-linear rejection transport technique 
  introduced by Stern et al.\ (1995),    
  based on the concept of a virtual cross-section
  (see Nelson, Hirayama, \& Rogers 1985; Kawrakow \& Rogers, 2001).
We note that in Stern et al.\ (1995), a constant density plasma is considered.  
We construct a more general algorithm 
  which integrates the mean-free-path over the spatially varying electron density 
  (Cullen 2001a, b).   
The Klein-Nishina formula is used in determing the scattering cross-section. 
At this scattering point, the momentum of the electron 
  is drawn from an isotropic Maxwellian distribution of the local temperature. 
The momentum vector is then Lorentz transformed 
  out of the rest frame of the local bulk flow to the rest frame of the white dwarf.  
A rejection algorithm is then used to decide whether to accept or reject the scattering event, 
  with an acceptance probability $\sigma_{\rm KN}V_{\rm rel}/2 \sigma_{\rm T} c$
  (Cullen 2001b), 
  where $\sigma_{\rm KN}$ is the Klein-Nishina cross-section, 
  $\sigma_{\rm T}$ is the Thomson cross-section, $c$ is the speed of light, and 
  $V_{\rm rel}$ is the velocity of the electron relative to the photon.    
For an accepted event, the energy and momentum changes are computed 
  using the scattering algorithm described in \citet{Pozdnyakov83}.

In each simulation, 
   the photons are tagged and followed until they escape from the accretion column. 
If a photon impacts with the white-dwarf surface after leaving the accretion column, 
  it is treated as absorbed and is discarded.
The remaining escape photons are binned to construct a spectrum.  
The most recent version of the Mersenne Twister random number generator 
  (Matsumoto \& Nishimura 1998) was used in our simulations. 
A full description of the numerical algorithm can be found in Cullen (2001a). 

\begin{figure*}
\centerline{\includegraphics[width=13.0truecm]{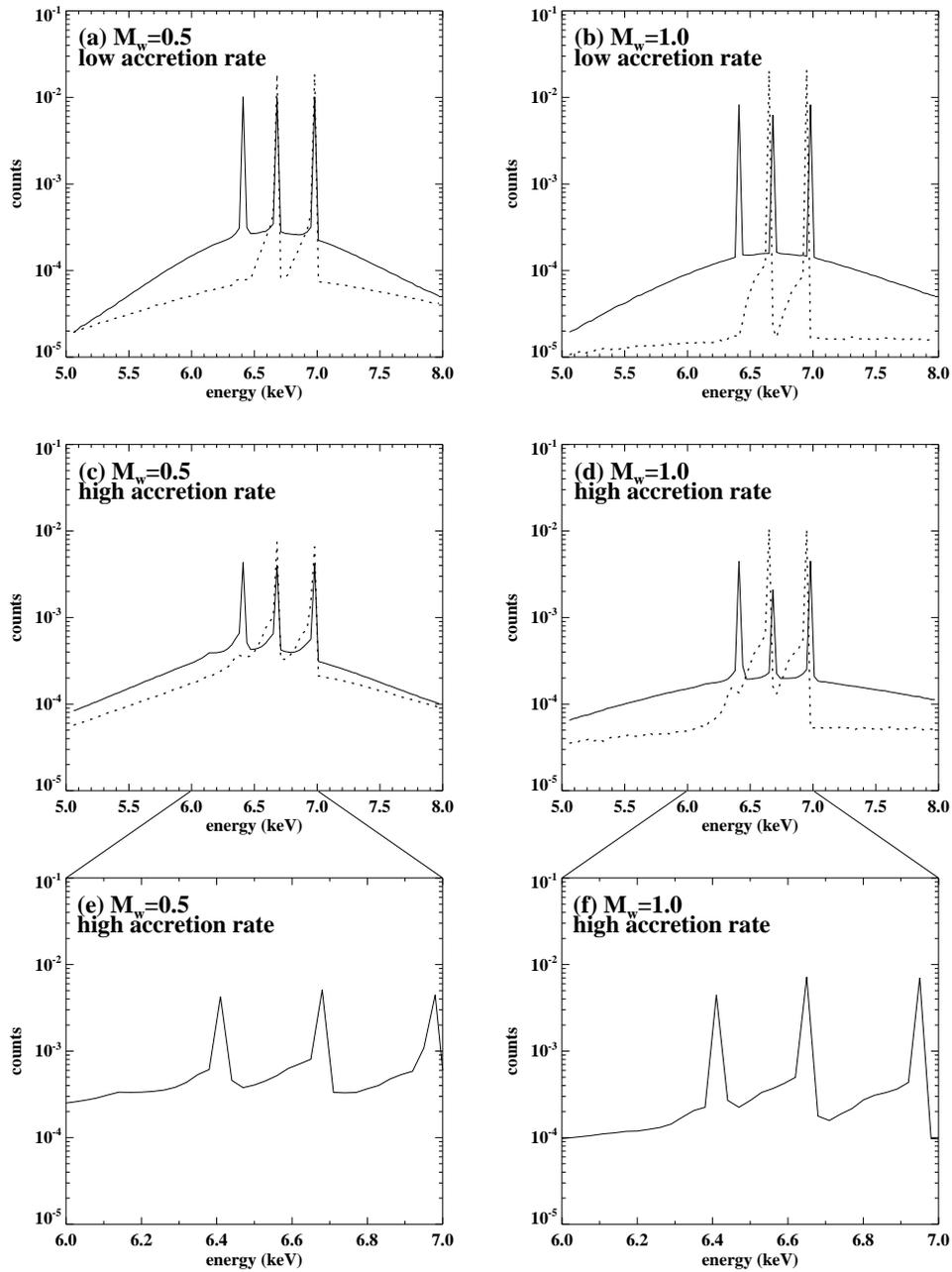}}
\caption{Simulated Fe K$\alpha$ emission lines scattered by electrons in the accretion column  
  of a white dwarf of mass 0.5 $M_\odot$ (left column) and 1.0 $M_\odot$ (right column), 
  and with a mass accretion rate of $1\,{\rm g\, cm}^{-2}{\rm s}^{-1}$ 
  and $10\,{\rm g\, cm}^{-2}{\rm s}^{-1}$.
Cyclotron cooling is set to zero.
The solid and dotted lines 
  correspond to cases for photon injection sites 
  near the white-dwarf surface and at the accretion shock, respectively.
The profiles in (e) and (f) are a close-up of the $6 - 7\,$keV region,
showing the 6.4~keV line emitted near the surface, 
  and the 6.675 and 6.97~keV lines emitted at the shock, under the same conditions as
those for (c) and (d).}
\end{figure*}

\section{Results and Discussion}

We simulated electron scattering off Fe~K$\alpha$ photons 
  with emission line central energies 6.4~keV, 6.675~keV, and 6.97~keV 
  using a total of $10^8$ photons.
We considered two white dwarf masses and radii: 
  $M_{\rm w} = 0.5 M_\odot$,
  $R_{\rm w} = 9.2 \times 10^8\,$cm, and
  $M_{\rm w} = 1.0 M_\odot$, 
  $R_{\rm w} = 5.5 \times 10^8\,$cm.
For each mass--radius, 
  we considered two different mass accretion rates: 
  $\dot m = 1\,{\rm g\, cm}^{-2}{\rm s}^{-1}$ and 
  $\dot m = 10\,{\rm g\, cm}^{-2}{\rm s}^{-1}$.
The 6.4~keV photons were injected 
  near the white dwarf surface,
  where the bulk speed of the accreting material goes to zero.
The 6.675 and 6.97~keV photons were each injected 
  at either the white dwarf surface or the shock, 
  where the bulk speed is equal to $v_{\rm ff}/4$.

Figure~2 shows the results 
  for when cyclotron cooling is negligible and bremsstrahlung cooling dominates 
  ($\epsilon_{\rm s} = 0$).
Figure~3 shows the results for when cyclotron emission is the dominant cooling process 
  (with $\epsilon_{\rm s}=100$).
(A more detailed investigation into the  line emissivity profile in the accretion column, 
  taking into account collisional ionisation, is deferred for future work.)
All the line profiles in Figure~2 show broadening effects due to electon (Compton) scattering. 
The broadening near the base of the profiles is due to multiple scatterings of photons 
  by the hot electrons in the post-shock region. 
This is most evident in all cases where photons are injected at the base of the accreting column 
 (denoted by solid line profiles in Fig.~2). 
As line photons emitted near the white-dwarf surface 
  have further to propagate through the accreting material before escaping the column, 
  the probability of scattering is larger 
  than that for photons emitted further above the white dwarf surface, near the shock. 
Line photons emitted near the shock (with profiles denoted by dotted lines in Fig.~2) 
  encounter fewer electrons in the post-shock region and thus,
  broadening near the base of the line profiles is weaker.   
The broad bump caused by multiple scatterings 
  can, however, be  washed out by the continuum emission 
  (which is emitted mostly at the dense base, see Wu 2000; Cropper et al. 2000),
  which also undergoes the same degree of scattering.   

Distortions due to Compton recoil by cold electrons can be seen more clearly 
  in the line profiles of photons emitted at the shock.
The signature of Compton recoil, a small shoulder red-ward of the line, 
  can be seen especially in the high-$\dot m$ cases, Figs.~2(c) and (d).
This is because the electron number density increases linearly with $\dot m$, 
  so the optical depth, and hence, the scattering probability, increases.
A comparison of Fig.~2(c) and (d) also indicates that the shoulder-like Compton recoil signature 
  is more conspicuous in high-mass mCVs, 
  if other parameters are kept constant in the model. 
The effects of Compton recoil scattering by cold pre-shock electrons 
  are less likely to be washed out by continuum emission  
  provided that the seed line photons are injected near the shock 
  (not the dense base of the post-shock region),
  and that the specific accretion rate $\dot m$ is sufficiently high, 
  as indicated in Figs.~2(d) and (f). 

\begin{figure}
\centerline{\includegraphics[width=7.0truecm]{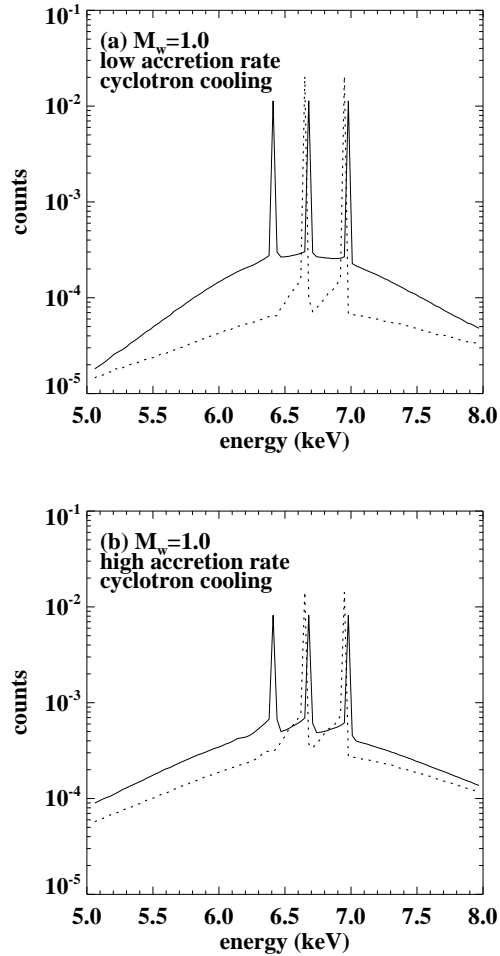}}
\caption{Simulated Fe K$\alpha$ emission lines scattered by electrons
in the  accretion column of a white dwarf
of mass 1.0 $M_\odot$, and mass accretion rate
$10\,{\rm g\, cm}^{-2}{\rm s}^{-1}$, with cooling in the post-shock region
dominated by cyclotron emission.
The solid and dotted lines are for photon injection sites near the white
dwarf surface and at the accretion shock, respectively.}
\end{figure}

Our simulations show 
  that the down-scattered photons of the H- and He-like Fe lines 
  emitted near the accretion shock can contaminate the neutral Fe~K$\alpha$ line  
  emitted near the white-dwarf surface.
This can be seen in Figs.~2(e) and (f), 
  which show close-ups of the line profiles over the energy range 6$-$7~keV, 
  with the 6.4~keV line emitted at the white-dwarf surface, 
  and the 6.675 and 6.97~keV lines both emitted at the shock.
The white-dwarf mass-radius and accretion rate are 
  the same as in Figures~2(c) and (d), as indicated. 

Figure~3 shows simulated line profiles with the same input parameters as those for Fig.~2(b) and (c), 
  but with the ratio of cyclotron to bremmstrahlung cooling ($\epsilon_{\rm s}$) set to 100.
Qualitatively, we expect additional cooling of the electrons in the post-shock region 
  to substantially change the thickness and the density structure of the post-shock region 
  and hence, the scattering optical depth. 
Moreover, it also alters the spatial Fe~K$\alpha$ line emissivity profiles in the region. 
Our simulations in Figure~3 indicate that this enhances the overall Compton scattering
effects,  particularly in high $\dot m$ mCVs.
Note, however, that the enhanced broadening near the base of the line profiles makes
the recoil signatures less discernable.
On the other hand, this also indicates that the Compton broadening of the base may be
less likely than in the case of no cyclotron cooling to be washed out by continuum
emissions and thus, is potentially more observable.
There is some tentative evidence for this effect in \textit{Chandra} High-Energy
Transmission Grating (HETG) spectrum
of the polar AM Her, particularly when compared to the HETG spectruum of AO Psc,
which is an IP and thus, has less cyclotron cooling (Hellier \& Mukai, 2004).
However, higher signal-to-noise observations are required to confirm this Compton
scattering effect.

We note that in some mCV X-ray observations, the Fe lines appear to have centre energies
slightly lower than 
  expected (Ishida 1991, see also Hellier et al., 1998; Hellier \& Mukai, 2004). 
In a typical observation a small red shoulder caused by recoil scattering of line photons 
  would not be easily resolved by instruments with insufficient spectral resolution. 
The resulting lines in the observed spectrum would thus appear to be slightly red-shifted. 
Thus, recoil scattering is a plausible interpretation of the observed red-shifts
of Fe lines in mCV X-ray spectra.  
Another possible interpretation, suggested by Hellier \& Mukai (2004) is that 
satellite lines associated with the H-like and He-like transitions are responsible
for the apparent line centre shifts.
These interpretations will be verifiable only when spectra with sufficiently high
spectral resolution and signal-to-noise ratio to resolve the fine substructures
in the line profiles have been obtained
(e.g. by \textit{Astro-E2} and other future X-ray satellites such as \textit{Xeus} and
\textit{Con-X}).

\section{Summary and Conclusions}

We performed Monte Carlo simulations of electron (Compton) scattering 
  of Fe K$\alpha$ lines in the accretion column of mCVs 
  using a non-linear algorithm which 
  takes into account velocity, density and temperature variations in the scattering region.  
Our results can be summarised as follow: 
(1) Recoil scattering by electrons in the pre-shock accretion flow 
     can create a small shoulder-like feature red-ward of the line.  
   This effect is prominent for photons emitted near the shock,
     but is not obvious for photons emitted from the dense base near the white-dwarf surface.  
   Moreover, the effect is most noticeable
     for systems with a high-mass ($\sim 1\,M_\odot$) white dwarf, and
     with a high mass accretion rate ($\sim 10\,{\rm g\, cm}^{-2}{\rm s}^{-1}$). 
(2) The 6.4~keV neutral Fe line can be contaminated by the photons 
     down-scattered from the higher energy H- and He-like lines.   
(3) Lines emitted from the dense base of the accretion column experience 
    multiple scatterings, and a very broad bump is formed at the base of the line
profiles. However, the broad bump is unlikely to be distinguishable from the continuum
emission, which is predominantly emitted from the dense base of the column and which
consequently suffers a similar degree of scattering.
The possible exception to this is when cyclotron cooling is sufficiently strong to
enhance the density structure in the accretion column, thereby enhancing the effects
of Compton scattering in the post-shock region, and the prominence of the resulting
broadening at the base of the line profiles.



\section*{Acknowledgments} 
ZK thanks the Australian Academy of Science for financial support,
and the Mullard Space Science Laboratory for their hospitality.
The authors also thank an anonymous referee whose comments helped
to improve the paper.


\end{document}